\documentstyle[twoside,fleqn,espcrc2,epsfig]{article}

\def\proof{\noindent{\sl Proof:}\kern0.6em}

\def\frac#1#2{\hbox{$#1\over#2$}}
\def\dual{\mathstrut^*\kern-0.1em}

\def\lvec#1{\setbox0=\hbox{$#1$}
    \setbox1=\hbox{$\scriptstyle\leftarrow$}
    #1\kern-\wd0\smash{
    \raise\ht0\hbox{$\raise1pt\hbox{$\scriptstyle\leftarrow$}$}}
    \kern-\wd1\kern\wd0}
\def\rvec#1{\setbox0=\hbox{$#1$}
    \setbox1=\hbox{$\scriptstyle\rightarrow$}
    #1\kern-\wd0\smash{
    \raise\ht0\hbox{$\raise1pt\hbox{$\scriptstyle\rightarrow$}$}}
    \kern-\wd1\kern\wd0}

\def\nabstar#1{\nabla\kern-0.5pt\smash{\raise 4.5pt\hbox{$\ast$}}
               \kern-4.5pt_{#1}}

\def\drvstar#1{\partial\kern-0.5pt\smash{\raise 4.5pt\hbox{$\ast$}}
               \kern-5.0pt_{#1}}

\def\rhoprime{\rho\kern1pt'}
\def\rhobar{\bar{\rho}}
\def\rhobarprime{\rhobar\kern1pt'}
\def\rhobartilde{\kern2pt\tilde{\kern-2pt\rhobar}}
\def\rhobartildeprime{\kern2pt\tilde{\kern-2pt\rhobar}\kern1pt'}

\def\zetabar{\bar{\zeta}}
\def\zetaprime{\zeta\kern1pt'}
\def\zetabarprime{\zetabar\kern1pt'}
\def\zetar{\zeta_{\raise-1pt\hbox{\sixrm R}}}
\def\zetabarr{\zetabar_{\raise-1pt\hbox{\sixrm R}}}

\def\phiimpr{\phi_{\kern0.5pt\hbox{\sixrm I}}}

\def\diracstar#1#2{
    \setbox0=\hbox{$\gamma$}\setbox1=\hbox{$\gamma_{#1}$}
    \gamma_{#1}\kern-\wd1\kern\wd0
    \smash{\raise4.5pt\hbox{$\scriptstyle#2$}}}

\def\ba{b_{\rm A}}

\def\bp{b_{\rm P}}

\def\ca{c_{\rm A}}

\def\f1{f_1}

\def\opprime#1{\setbox0=\hbox{${\cal O}$}\setbox1=\hbox{${\cal O}_{\rm #1}$}
    {\cal O}_{\rm #1}\kern-\wd1\kern\wd0
    \smash{\raise4.5pt\hbox{\kern1pt$\scriptstyle\prime$}}\kern1pt}

\def\ophatprime#1{\setbox0=\hbox{$\widehat{\cal O}$}
    \setbox1=\hbox{$\widehat{\cal O}_{\rm #1}$}
    \widehat{\cal O}_{\rm #1}\kern-\wd1\kern\wd0
    \smash{\raise4.5pt\hbox{\kern1pt$\scriptstyle\prime$}}\kern1pt}

\def\bopprime#1{\setbox0=\hbox{${\cal O}$}\setbox1=\hbox{${\cal O}_{\rm #1}$}
    {\cal L}_{\rm #1}\kern-\wd1\kern\wd0
    \smash{\raise4.5pt\hbox{\kern1pt$\scriptstyle\prime$}}\kern1pt}

\def\blagprime#1{\setbox0=\hbox{${\cal B}$}\setbox1=\hbox{${\cal B}_{#1}$}
    {\cal B}_{#1}\kern-\wd1\kern\wd0
    \smash{\raise5.2pt\hbox{\kern1pt$\scriptstyle\prime$}}\kern1pt}

\def\mq{m_{\rm q}}

\def\msbar{{\rm \overline{MS\kern-0.05em}\kern0.05em}}

\begin{document}
\title{ \vspace{-4.0cm}
       \rightline{\normalsize CERN-TH/2001-269}
       \vspace{-0.1cm}
       \rightline{\normalsize HU-EP-01/37}
       \vspace{-0.1cm}
       \rightline{\normalsize October 2001}
       \vspace{2.0cm}
The charm quark's mass in quenched QCD\thanks{ 
presented at the ${\rm XIX}$ International Symposium on 
Lattice Field Theory ``Lattice 2001'', 
August 19 -- 24, 2001, Berlin, Germany}}

\author{Juri Rolf\address{Institut f\"ur Physik, Humboldt-Universit\"at zu 
Berlin, Invalidenstr. 110, D-10115 Berlin, Germany}
 and Stefan Sint\address{CERN, Theory Division, CH-1211 Geneva 23, Switzerland}
}

\begin{abstract}
We present our  preliminary result for the charmed quark mass,
which follows from taking the $D_s$ and $K$ meson masses 
from experiment and $r_0=0.5\,{\rm fm}$ (or, equivalently 
$F_K=160\,{\rm MeV}$) to set the scale. 
For the renormalization group invariant quark mass we obtain
$M_c = 1684(64)\, {\rm MeV}$, which translates to
$m_c(m_c)= 1314 (40)(20)(7)\,{\rm MeV}$ for the running mass in the
$\overline{\rm MS}$ scheme. Renormalization is treated non-perturbatively,
and the continuum limit has been taken, so that the only 
uncontrolled systematic error consists in the use of the quenched 
approximation.
\end{abstract}
\maketitle

\section{INTRODUCTION}

Quark masses are among the fundamental parameters 
of the Standard Model, and yet not directly accessible to
experimental measurement. Their determination 
in terms of experimental quantities requires  
a good quantitative control of the non-perturbative dynamics
of QCD, which may be achieved by numerical
simulations of the lattice regularized theory.
Extending previous work on the $\Lambda$-parameter 
and the strange quark mass in quenched 
QCD~\cite{Capitani:1999mq,Garden:2000fg}
we report here on a precise determination 
of the charm quark mass, taking the experimental
value of the  $D_s$ meson mass as essential new input.
For previous work on the charm quark mass we refer to
two recent papers~\cite{Kuhn:2001dm}
and the references given there.

\section{STRATEGY}

We use O($a$) improved Wilson quarks with the Wilson gauge
action. All necessary renormalization constants and
improvement coefficients are known with non-perturbative accuracy
in the relevant range of bare couplings.
We refer to~\cite{Sint:2001vc} for a review and further references.

\subsection{Setting the parameters}

In quenched QCD it is convenient to measure dimensionful quantities
in units of the scale $r_0$~\cite{Sommer:1994ce}, since
$r_0/a$ has been determined very precisely for a 
large range of bare couplings~\cite{Guagnelli:1998ud,Necco:2001xg}.
To obtain the scale in physical units we always use
$r_0=0.5\,\,{\rm fm}$, which is roughly equivalent to
setting the scale with $F_K=160\,\,{\rm MeV}$~\cite{Garden:2000fg}.
The renormalization group invariant (RGI) 
strange quark mass has been obtained in~\cite{Garden:2000fg},
\begin{equation}
  r_0M_s = 0.348(13) \Rightarrow M_s = 138(6)\,\,{\rm MeV},
  \label{Mstrange}
\end{equation}
taking $m_K$ from experiment and the mass ratio 
\begin{equation}
  M_s/\hat{M}= 24.5 \pm 1.5,\quad  \hat{M}=\frac12(M_u+M_d),
\end{equation}
from chiral perturbation theory~\cite{Leutwyler:1996qg}.
Up to intrinsic O($a^2$) ambiguities, the connection 
between $M_s$ and the bare strange quark mass 
can be established by combining
the results of refs.~\cite{Capitani:1999mq,Guagnelli:2001jw},
so that no tuning is required to satisfy the renormalization
condition~(\ref{Mstrange}).
The bare charm quark mass may then be fixed by matching 
directly the experimental result 
$m_{D_s}=1969\,{\rm MeV}$~\cite{Groom:2000in}, 
which translates to $r_0 m_{D_s}=4.98$.
This is justified since electroweak effects on $D$ meson masses 
are expected to be small on the scale of the expected
statistical errors.

\subsection{Some technical details}

The simulations were done at four values of $\beta$,
$6.0\leq\beta\leq6.45$, which correspond to lattice spacings
in the range $0.05-0.1\,{\rm fm}$. Spatial lattice volumes
ranged from $16^3$ to $32^3$, and the Euclidean time extent
from $40$ to $80$. The linear extent of the spatial volume was never less
than $1.5\,\,{\rm fm}$ in physical units, so that
finite volume effects on the $D_s$ meson mass can be safely neglected. 
Quark propagators were computed for two mass parameters in the 
strange quark region, and three values around the expected charm quark mass. 
To compute meson masses we used correlation functions
derived from the Schr\"odinger functional (SF)~\cite{Guagnelli:1999zf}. 
In the pseudoscalar channel these are given by
\begin{equation}
  f_{\rm X}(x_0) = -\frac12a^6\sum_{\bf y,z}\left\langle 
                     \bar{c}(x)\Gamma^{}_{\rm X} s(x) 
                     \zetabar_s({\bf y})\gamma_5\zeta_c({\bf z})
                                     \right\rangle, 
\end{equation}
where $\Gamma^{}_{\rm X}$ stands for 
$\gamma_0\gamma_5,\,\gamma_5$ for ${\rm X}={\rm A,P}$ respectively.
The pseudoscalar meson mass is then obtained by
looking for a plateau in the effective masses,
\begin{equation}
   am^{\rm X}_{\rm eff}(x_0+\frac12 a) = 
  \ln\left\{f_{\rm X}(x_0)/f_{\rm X}(x_0+a)\right\},
\end{equation}
as a function of Euclidean time $x_0$.
Autocorrelation times were found to be
completely negligible, so that the produced O(100) 
configurations at each $\beta$-value are statistically independent. 
An example for an effective mass plot is given in fig.~1. Excited
states are expected to contribute significantly
when either $x_0$ or $T-x_0$ becomes small. It is 
possible to estimate these contributions and thereby quantify
the resulting deviation from the true plateau value.
Requiring this effect to be smaller than $0.5\%$,
the plateau region in fig.~1 was identified with the interval
$25 \leq x_0/a \leq 35$; the meson mass was then
estimated by averaging the effective mass in this interval.
\begin{figure}[htb]
  \epsfig{file=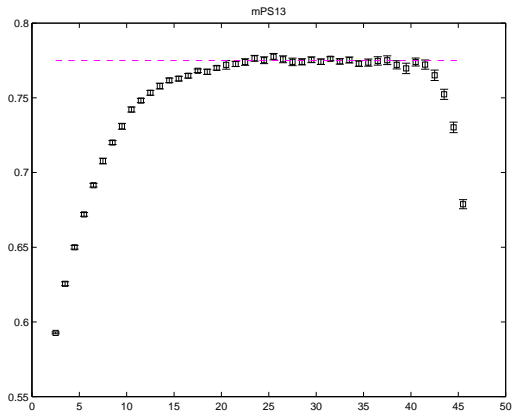,width=7cm}
  \vskip -5ex
  \caption{Effective pseudoscalar mass in
  lattice units vs.~$x_0/a$ at $\beta=6.1$.}
  \vskip -4ex
  \label{plateau}
\end{figure}
\subsection{Bare and renormalized quark masses}

Given the pseudoscalar masses at the chosen parameters,
an interpolation in some bare charm quark 
mass is required to match the experimental $D_s$ meson mass. 
Besides the bare mass $m_0$ we also consider 
two alternative definitions of the bare charm quark mass, which
derive from the PCAC relation. First we use the 
heavy--light axial current and density to define, 
\begin{equation}
   m_{cs}(x_0) = {{\tilde\partial_0 f_{\rm A}(x_0)+
   \ca a\partial^\ast_0\partial_0 f_{\rm P}(x_0)}
   \over{2f_{\rm P}(x_0)}},
\end{equation} 
which upon renormalization yields the average charm and strange
quark mass. With our parameter choices
yet another bare charm quark mass can be obtained from 
the PCAC relation involving a hypothetical strange quark,
assumed to be mass degenerate with the charm quark. 
We denote this PCAC mass as $m_{cc}$ but one should bear in mind that
its definition does not involve flavour singlet operators.

The renormalized and O($a$) improved charm quark mass $m_{\rm R}$ 
can now be obtained in various ways. Choosing the SF scheme 
we first obtain 
\begin{equation}
  m_{\rm R}=Z_{\rm P}^{-1}Z \mq(1+b_{\rm m}a\mq),
\end{equation}
where $\mq$ is the subtracted bare charm quark mass.
From the PCAC mass $m_{cs}$, one gets
\begin{equation}
 m_{\rm R}+m_{{\rm R},s} = Z_{\rm P}^{-1}Z_{\rm A}
  \left\{1+(\ba-\bp)a\overline{m}_{\rm q}\right\} 2m_{cs}, 
 \label{PCACmixed}
\end{equation}
where $\overline{m}_{\rm q}$ 
denotes the average of the bare subtracted charm and strange quark masses,
and we have assumed $x_0=T/2$, which is 
kept approximately constant in physical units.
Finally, we consider
\begin{equation}
  m_{\rm R} = Z_{\rm P}^{-1}Z_{\rm A}
  \left\{1+(\ba-\bp)am_{\rm q}\right\} m_{cc}, 
\end{equation}
and convert to the RGI quark mass using
the flavour-independent ratio~\cite{Capitani:1999mq}
\begin{equation}
   M/m_{\rm R} = 1.157(15).
  \label{RGIratio}
\end{equation}

\section{RESULTS AND CONCLUSIONS}

After subtraction of the strange quark 
mass~(\ref{Mstrange}) in eq.~(\ref{PCACmixed})
we have three definitions of $r_0M_c$ which should coincide 
up to terms of O($a^2$). In each case 
we extrapolate to the continuum with an
ansatz of the form
\begin{equation}
  r_0M_c = c_0 + c_1(a/r_0)^2,
\end{equation}
excluding the data point at the coarsest lattice spacing ($\beta=6.0$).
As can be seen in fig.~2, the results are nicely compatible 
with each other in the continuum limit, 
although the differences at finite cutoff are quite large.
As our best (preliminary) estimate we quote
\begin{equation}
  r_0M_c = 4.26(16) \Rightarrow M_c=1684(64)\,{\rm MeV}.
\end{equation}
Using the four-loop anomalous dimension and $\beta$-function
in the $\overline{\rm MS}$ scheme~\cite{vanRitbergen:1997va},
and  $r_0\Lambda_{\overline{\rm MS}}=0.586(48)$~\cite{Necco:2001xg},
we obtain
\begin{equation}
  m_c(m_c)=1314(40)(20)(7)\,{\rm MeV},
\end{equation}
where the first error stems from the RGI quark mass, the second is
induced by the error on $\Lambda_{\overline{\rm MS}}$, 
while the last number is the difference
between three- and four-loop perturbative evolution down to the
charm scale.

We conclude that precise results can be obtained for
charmed observables using the same techniques 
as for the light quarks, provided that the continuum extrapolation
is performed. We emphasize that we have used the quenched approximation,
which is known to lead to inconsistencies in the  hadron spectrum
at the $10\%$ level. However, the present result 
constitutes a significant improvement over previous lattice estimates
and will be a useful reference point for more realistic
calculations involving dynamical quarks. For the time being we
speculate that the mass ratio  $M_c/M_s = 4.26(16)/0.348(13)
\approx 12.2\pm 1.0$ may be less sensitive to quenched 
scale ambiguities than the masses themselves.
\begin{figure}[htb]
 \epsfig{file=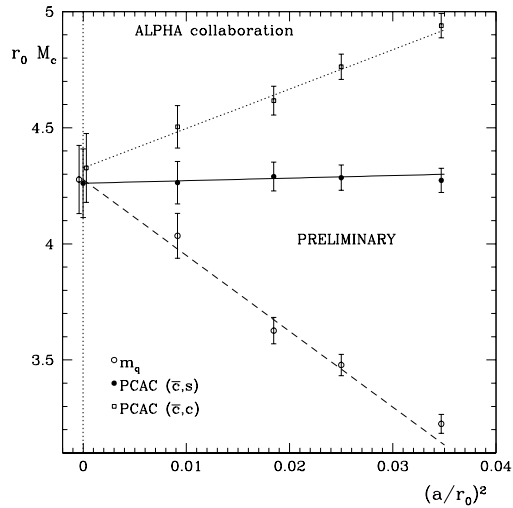,width=7cm}
 \vskip -5ex
 \caption{Continuum extrapolation of the RGI charm quark masses.}
 \vskip -4ex
 \label{contextrap}
\end{figure}

\vskip 2ex

This work is part of the ALPHA collaboration research programme,
and partially supported by the European Community under the 
grant HPRN-CT-2000-00145 Hadrons/Lattice QCD.
Simulations were carried out on machines of the APE100 and APE1000
series at DESY-Zeuthen. We thank the staff at the computer centre
for their help, and P.~Ball, R.~Sommer, H.~Wittig and U.~Wolff
for useful discussions.

\vfill
\eject

\end{document}